\documentclass[preprint,showpacs,preprintnumbers,amsmath,amssymb,nofootinbib]{revtex4-1}

\usepackage{graphicx}
\usepackage{dcolumn}
\usepackage{bm}

\newcommand{\beq}{\begin{equation}}
\newcommand{\eeq}{\end{equation}}
\newcommand{\beqy}{\begin{eqnarray}}
\newcommand{\eeqy}{\end{eqnarray}}
\newcommand{\beqyn}{\begin{eqnarray*}}
\newcommand{\eeqyn}{\end{eqnarray*}}
\newcommand{\nl}{\newline}
\newcommand{\nn}{\nonumber}

\newcommand{\bc}{\begin{center}}
\newcommand{\ec}{\end{center}}
\newcommand{\bmin}{\begin{minipage}}
\newcommand{\emin}{\end{minipage}}

\begin{document}

\title{New relation between transverse angular momentum and generalized parton distributions}

\author{Elliot Leader}
 \email{e.leader@imperial.ac.uk}
\affiliation{Blackett laboratory \\Imperial College London \\ Prince Consort Road\\ London SW7 2AZ, UK}

\date{\today}

\begin{abstract}
I derive a rigorous relation between the expectation value of the transverse component of the Bellinfante version of the angular momentum $\langle \, J_T^{bel} \, \rangle $ of a quark in a transversely polarized nucleon in terms of the Generalized Parton Distributions $H$ and $E$, namely
\beq \langle \, J_T^{bel}(\textrm{quark}) \, \rangle =\frac{1}{2M }\, \left[ P_0\,\int_{-1}^{1} dx x E_q(x,0,0) + M \, \int_{-1}^{1} dx x H_q(x,0,0) \right] \nn  \eeq
where $P_0$ is the energy of the nucleon and where ``quark" implies the sum of quark and antiquark of a given flavor. A similar relation holds for gluons. The result is remarkably similar to Ji's relation for the case of longitudinal polarization, whose content we spell out precisely, and which  has proved extremely interesting in testing models, in comparing different definitions of quark and gluon angular momentum, and in comparing with lattice calculations.The transverse relation can offer additional insights into all of these, by extending them to the new domain of transversely polarized and moving nucleons.
\end{abstract}

\maketitle

Some time ago Ji \cite{Ji:1997pf, *Ji:1996ek, Ji:1996nm}, using the Bellinfante version of the angular momentum, derived an expression for the expectation value of the longitudinal component of the quark angular momentum in a nucleon $\langle \, J_L^{bel}(\textrm{quark}) \, \rangle $ in terms of the Generalized Parton Densities (GPDs) $H$ and $E$, namely
\beq \label{Ji} \langle \, J_L^{bel}(\textrm{quark}) \, \rangle =\frac{1}{2 }\, \left[ \int_{-1}^{1} dx x E_q(x,0,0) +  \int_{-1}^{1} dx x H_q(x,0,0) \right] \eeq
where here and in the following ``quark" signifies the contribution of a \emph{quark plus antiquark of a given flavor}. Often this relation is misleadingly written without the label specifying the longitudinal component and without the label Bellinfante, but more importantly, it is often forgotten that since the quark angular momentum itself is not conserved, its matrix elements will   depend on the renormalization scale $\mu$. Thus this equation and \emph{all those to follow } should, strictly speaking, carry a label $\mu$.

I shall show that a very similar relation holds for the transverse component of the quark angular momentum, i.e.
\beq \label{J_T} \langle \, J_T^{bel}(\textrm{quark}) \, \rangle =\frac{1}{2M }\, \left[ P_0\,\int_{-1}^{1} dx x E_q(x,0,0) + M \, \int_{-1}^{1} dx x H_q(x,0,0). \right]   \eeq
The, at first sight, unexpected appearance of the nucleon energy $P_0$ will be discussed later. Note that  neither of these relations is really a \emph{sum rule} since, while we might be able to measure $H$ from deep inelastic lepton-hadron scattering, and $E$ from deeply virtual Compton scattering, we do not know how to measure $J_L^{bel}$ or $J_T^{bel}$ experimentally. Nonetheless they are important relations for testing models of the nucleon and comparing with lattice calculations.
In the following, to avoid controversy \cite{Leader:2011za},   I will utilize the Bellinfante definition of angular momentum, as was done by Ji.
The derivation is based on a comparison of the defining expression for the GPDs and the  expression for the angular momentum in terms of the matrix elements of the quark and gluon energy momentum tensors $t^{\mu\nu}_{q,\, bel} (0)$  and $t^{\mu\nu}_{G,\, bel} (0)$. \nl
 The connection between these matrix elements and the angular momentum involves divergent integrals, which have to be treated carefully using wave packets, as was done correctly for \emph{arbitrary} components of $\bm{J}$ for the first time by Bakker, Leader and Trueman  (BLT) \cite{Bakker:2004ib}, and for this reason I shall use their notation for the scalar functions that appear in the matrix element of $t^{\mu\nu}_{q, \, bel}(0)$. One has

\beqy \label{tmunu} \langle \, P',S' \, |\, t^{\mu\nu}_{q, \, bel}(0)\,  \, | \, P, S \, \rangle &=&[\bar{u}'\gamma^\mu u \, \bar{P}^\nu + (\mu \leftrightarrow \nu)]\mathbb{D}_q(\Delta^2)/2 \nn \\
&& \hspace{-2cm}+\frac{1}{2} \left[\frac{i\Delta_\rho}{2M}\, \bar{u}' \sigma^{\mu\rho} u \, \bar{P}^\nu + (\mu \leftrightarrow \nu) \right] [2\mathbb{S}_q(\Delta^2) - \mathbb{D}_q(\Delta^2)]\nn \\
 && \hspace{-4cm}+ \frac{\bar{u}' u}{2M}\left[\frac{1}{2}[\mathbb{G}_q(\Delta^2)-\mathbb{H}_q(\Delta^2)](\Delta^\mu \Delta^\nu - \Delta^2 g^{\mu\nu}) +M^2 \mathbb{R}_q(\Delta^2)g^{\mu\nu}\right]
 \eeqy
 where
 \beq   \bar{P}=\frac{1}{2}(P + P')  \qquad \Delta = P'-P \qquad  u\equiv u(P,S)  \qquad u'\equiv u(P',S') \eeq and
 the spinors are normalized to $\bar{u}u=2M$.
 Note that the term $M^2 \mathbb{R} g^{\mu\nu}$ is only allowed because we are dealing with  a non-conserved density. \nl
 In the standard notation (see e.g. the review of Diehl \cite{Diehl:2003ny}) the GPDs, for a nucleon moving in the positive $Z$ direction, are defined by
\beqy \label{GPD} &&\frac{1}{2}\int \frac{dz^-}{2\pi}\, e^{i x \bar{P}^+ z^-} \,\langle \, P' \,S' \,| \,\bar{\psi}(-z^-/2)\, \gamma^+ \, W \,\psi (z^-/2) \, | \, P \, S \,\rangle \nn \\
&=& \frac{1}{2 \bar{P}^+}\left\{ [\bar{u}(P')\gamma^+ u(P)] H_q(x, \xi, t) + \left[\frac{i\Delta_\rho}{2M}\bar{u}(P') \sigma^{+ \rho} u(P) \right] E_q(x,\xi, t) \right\} \eeqy
where
\beq  t= \Delta^2 \qquad \Delta^+= -2\xi\bar{P}^+ \eeq
and we are using the standard definition of the $\pm $ components of a vector i.e.
\beq v^{\pm} = \frac{1}{\sqrt{2}}(v_0 + v_z).  \eeq
 W is the Wilson line operator
\beq \label{Wilson} W\equiv W[-z^-/2 \, , \, z^-/2]= \mathcal{P} \,\exp \{ig \int_{-z^-/2}^{z^-/2} \,dz'\,  A^+_a(z'n)\, t^a \}, \eeq
 a matrix in colour space, with
\beq n=\frac{1}{\sqrt{2}}(1,0,0,-1). \eeq

Integrating over $x$ and after some manipulation
 Eq.~(\ref{GPD}) yields
 \beqy \label{intGPD}&& \frac{1}{2 \bar{P}^+}  \left\{ [\bar{u}'\gamma^+ u]\int dx x H_q(x, \xi, t) +  \left[\frac{i\Delta_\rho}{2M} \, \bar{u}' \sigma^{+ \rho} u \right] \int dx x E_q(x,\xi, t) \right\}\nn \\
 && = \frac{i}{4(P^+)^2}\langle \, P' \,S' \,| \,\bar{\psi}(0) \, \gamma^+ \, \overleftrightarrow{D}^+ \,\psi (0) \, | \, P \, S \,\rangle = \frac{1}{2(P^+)^2} \langle \, P',S' \, |\, t^{+ \, +}_{q, \, bel}(0)\,  \, | \, P,S \, \rangle \eeqy
 where
 \beq \overleftrightarrow{D}^+ = \overrightarrow{\partial}^+ - \overleftarrow{\partial}^+ -2ig A^+(0). \eeq
 From Eq.~(\ref{tmunu}), remembering that $g^{++}=0$ and that $\Delta^+= -2\xi \bar{P}^+$, one obtains
 \beqy \label{tplusplus}  \langle \, P',S' \, |\, t^{+ \, +}_{q, \, bel}(0)\,  \, | \, P,S \, \rangle &= &[\bar{u}'\gamma^+ u \, \bar{P}^+ ][\mathbb{D}_q(\Delta^2) + \xi^2 ( \mathbb{G}_q(\Delta^2)-\mathbb{H}_q(\Delta^2)) ]\nn \\
  && \hspace{-4cm} + \left[\frac{i\Delta_\rho}{2M}\, \bar{u}' \sigma^{+\rho}u \, \bar{P}^+  \right] [  2 \,\mathbb{S}_q(\Delta^2)- \mathbb{D}_q(\Delta^2) - \xi^2 ( \mathbb{G}_q(\Delta^2)-\mathbb{H}_q(\Delta^2))]. \eeqy
Upon taking the limit $\Delta \rightarrow 0 $, Eqs.~(\ref{intGPD}, \ref{tplusplus})  yield

  \beq \label{Hsum} \int_{-1}^{1} dx x H_q(x,0,0) = \mathbb{D}_q \eeq
  \beq \label{Esum} \int_{-1}^{1} dx x E_q(x,0,0)= (  2 \, \mathbb{S}_q -\mathbb{D}_q )\eeq
  and consequently
  \beq \label{HEsum} \frac{1}{2}\left[\int_{-1}^{1} dx x H_q(x,0,0) + \int_{-1}^{1} dx x E_q(x,0,0) \right] =  \mathbb{S}_q. \eeq

For the case of a \emph{longitudinally} polarized  nucleon moving  in the $Z$ direction BLT \cite{Bakker:2004ib} proved that $\mathbb{S}$ measures the expectation value of the $Z$-component of $\bm{J}$. Hence Eq.~(\ref{HEsum}) can be written
\beq \label{Jzsum} \frac{1}{2}   \int_{-1}^{1} dx x [H_q(x,0,0) +E_q(x,0,0)]=  \langle \, J_L^{bel}(\textrm{quark}) \, \rangle \eeq
which is the relation first derived by Ji \cite{Ji:1996nm}. \nl
Now it is well known that $\int_{-1}^{1} dx x H_q(x,0,0)$ measures the fraction of the nucleon's momentum carried by quarks and antiquarks of a given flavour, so that
 adding the gluon contribution\footnote{For gluons the integrals run from $0$ to $1$.}
\beq \label{Hfrac} \sum_{flavours} \int_{-1}^{1} dx x H_q(x,0,0) + \int_{0}^{1} dx x H_G(x,0,0) =1.\eeq
Hence, summing Eq.~(\ref{Jzsum}) over flavors and adding the analogous equation for gluons, one obtains
\beqy \label{qG} \frac{1}{2} + \sum_{flavors} \int_{-1}^{1} dx x E_q(x,0,0) + \int_{0}^{1} dx x E_G(x,0,0)&=& \sum_{flavours}\langle \, J_L^{bel}(\textrm{quark}) \, \rangle + \langle \, J_L^{bel}(\textrm{gluon}) \, \rangle \nn \\
&=& \frac{1}{2} \eeqy
so that
\beq \label{EqG} \sum_{flavors} \int_{-1}^{1} dx x E_q(x,0,0) + \int_{0}^{1} dx x E_G(x,0,0) =0. \eeq
This fundamental sum rule has wide ramifications and can be shown to correspond to the vanishing of the nucleon's anomalous gravitomagnetic moment \cite{Brodsky:2000ii, Teryaev:1999su}.

For the case of a \emph{transversely} polarized nucleon, moving along the positive $Z$ axis, it follows from BLT that
\beq  \label{Jx} \langle \, J_x^{bel}(\textrm{quark}) \, \rangle =\frac{1}{2M }\, \left[ (P_0\,(2 \,\mathbb{S}_{q}- \mathbb{D}_{q}) + M \, \mathbb{D}_{q} \right]  \eeq
Substituting Eqs.~(\ref{Hsum}, \ref{Esum}) we obtain the new result

\beq  \label{JT} \langle \, J_T^{bel}(\textrm{quark}) \, \rangle = \frac{1}{2M }\, \left[ P_0\,\int_{-1}^{1} dx x E_q(x,0,0) + M \, \int_{-1}^{1} dx x H_q(x,0,0) \right]  \eeq
where $P_0$ is the energy of the nucleon. \nl
The factor $P_0$ may seem unintuitive. However if we go the rest frame Eq.~(\ref{JT}) reduces to the Ji result Eq.~(\ref{Jzsum}), as it should, since in the rest frame there is no distinction between $X$ and $Z$ directions. Moreover, for a  classical relativistic system of particles, if one calculates the orbital angular momentum about the center of inertia for the system at rest, and then boosts the system one finds that the transverse angular momentum grows like $P_0$ \cite{Landau:1951}. Finally, if one sums Eq.~(\ref{JT}) over flavors and adds the analogous gluon equation, one finds that the term proportional to $P_0$ disappears, as it ought to, as a consequence of Eq.~(\ref{EqG}), and using Eq.~(\ref{Hfrac}), one obtains the correct result for a transversely polarized nucleon
\beq \label{sumJT} \sum_{flavors}\langle \, J_T^{bel}(\textrm{quark}) \, \rangle + \langle \, J_T^{bel}(\textrm{gluon}) \, \rangle = \frac{1}{2}. \eeq
The relation Eq~(\ref{JT}) can be used to test model results and also lattice calculations, since it is possible to treat a moving nucleon on a lattice.  \nl
Now BLT derived a sum rule for the \emph{total} angular momentum of a transversely polarized nucleon, namely
\beq \label{BLTTr} \frac{1}{2} = \frac{1}{2}\, \sum_{\textrm{flavours}}\, \int dx \, [ \Delta _T q (x) + \Delta _T \bar{q} (x)]
+ \sum_{q, \, \bar q, \, G }\langle L_T \rangle \eeq
where  $\Delta_T q(x)\equiv h_1(x)$ is the quark transversity distribution \footnote{ Note that the sum of quark and antiquark transversity densities does not correspond to the hadronic matrix element of a local operator and is unrelated to the chiral-odd tensor charge of the nucleon}. \nl
In this context it is important to realize that the quark part of Eq.~(\ref{BLTTr}) i.e.
\beq \label{qpart} \frac{1}{2}\, \sum_{\textrm{flavours}}\, \int dx \, [ \Delta _T q (x) + \Delta _T \bar{q} (x)]
+ \sum_{q, \, \bar q}\langle L_T \rangle \eeq
cannot be identified with $\langle \, J_T^{bel}(\textrm{quark}) \, \rangle $ in Eq.~(\ref{JT}). The reason is the following. While for the \emph{total} angular momentum there is no difference between Bellinfante and canonical angular momentum, i.e.
\beq \langle \, J_T^{bel}(\textrm{total}) \, \rangle = \langle \, J_T^{can}(\textrm{total}) \, \rangle \eeq
this is not true for the separate quark and gluon pieces, i.e.
\beq \langle \, J_T^{bel}(\textrm{quark}) \, \rangle \neq  \langle \, J_T^{can}(\textrm{quark}) \, \rangle \eeq
and in deriving Eq.~(\ref{BLTTr}) BLT used the property that $\bm{J}$ is the generator of rotations. As explained in detail in \cite{Leader:2011za} it is the canonical versions of the operators, $\bm{J}_{can}$, which are the generators of rotations. Thus the expression in Eq.~(\ref{qpart}) corresponds to $\langle \, J_T^{can}(\textrm{quark}) \, \rangle$ and should not be confused with $\langle \, J_T^{bel}(\textrm{quark}) \, \rangle$. \nl
Consider now the Bellinfante quark angular momentum operator which consists of a spin and an orbital term ( the spin term is the same in Bellinfante and canonical versions)
\beq \label{S+O} \bm{J}^{bel}(\textrm{quark}) = \bm{S}(\textrm{quark}) + \bm{L}^{bel}(\textrm{quark}) \eeq
where
\beq \label{S} \bm{S}(\textrm{quark}) = \int d^3x \, \bar{\psi}(x) \bm{\gamma} \gamma_5 \psi(x). \eeq
For a nucleon in a state with covariant spin vector $\mathcal{S}$
\beq \label{cov} \mathcal{S}^\mu = ( \mathcal{S}_0, \bm{\mathcal{S}} ) = \left( \frac{\bm{P}\cdot \bm{s}}{M}, \, \, \bm{s} \, + \frac{\bm{P}\cdot \bm{s}}{M(P_0 + M}\, \bm{P} \right) \eeq
where $\bm{s}$ is the rest frame spin vector, one has, for the expectation value 
\beqy \label{exS} \langle \bm{S}(\textrm{quark}) \, \rangle &=& \frac{1}{2P_0}\, \langle P, \, \mathcal{S}\, |\, \bar{\psi}(0) \bm{\gamma} \gamma_5 \psi(0)\,| \, P, \, \mathcal{S} \, \rangle \nn \\
= \frac{a_0^f M}{P_0} \,  \bm{\mathcal{S}} &= &   \frac{M}{P_0}\, \bm{\mathcal{S}} \,\frac{1}{2} \, \int_0^1 dx \,[ \Delta q(x) + \Delta \bar{q}(x)]_{\overline{MS}}
 \eeqy
 where $a_0^f$ is the contribution to the nucleon axial charge $a_0$ from a quark plus antiquark of flavor $f$. Its expression in terms of the longitudinal polarized parton densities is scheme dependent \cite{Leader:1999vs}.
For longitudinal polarization $ \mathcal{S}_z = P_0/M $, yielding the usual result
\beq \label{long} \langle S_L(\textrm{quark}) \, \rangle =  \frac{1}{2} \, \int_0^1 dx \,[ \Delta q(x) + \Delta \bar{q}(x)]_{\overline{MS}} \eeq
For transverse polarization, say in the $X$ direction, $ \mathcal{S}_x = 1 $ so that
\beq \label{trans} \langle S_T(\textrm{quark}) \, \rangle =  \frac{M}{2P_0} \, \int_0^1 dx \,[ \Delta q(x) + \Delta \bar{q}(x)]_{\overline{MS}}. \eeq
 In the case of longitudinal polarization, use of  the Ji relation to estimate
 \beqy \label{Jilong}  \langle \, L_z^{bel}(\textrm{quark}) \, \rangle &=& \frac{1}{2 }\, \left[ \int_{-1}^{1} dx x E_q(x,0,0) +  \int_{-1}^{1} dx x H_q(x,0,0) \right]\nn \\
 & - &\frac{1}{2} \, \int_0^1 dx \,[ \Delta q(x) + \Delta \bar{q}(x)]_{\overline{MS}} \eeqy
has proved extremely interesting in testing models, in comparing different definitions of quark and gluon angular momentum, and in comparing with lattice calculations. The transverse relation can offer additional insights into \textbf{all of these}, by extending them to the new domain of transversely polarized and moving nucleons.  In particular it would be very interesting to see how models or lattice calculations reproduce the $P_0$-dependence\footnote{As expected Eq.~(\ref{Ltran}) agrees with Eq.~(\ref{Jilong})in the rest frame where $P_0=M$. } in
\beqy \label{Ltran}  \langle \, L_x^{bel}(\textrm{quark}) \, \rangle &= & \frac{1}{2M }\, \left[ P_0\,\int_{-1}^{1} dx x E_q(x,0,0) + M \, \int_{-1}^{1} dx x H_q(x,0,0) \right]\nn \\
  &  - &  \frac{M}{2P_0} \, \int_0^1 dx \,[ \Delta q(x) + \Delta \bar{q}(x)]_{\overline{MS}}.   \eeqy
In summary I have derived a rigorous relation between the expectation value of the Bellinfante version of the transverse angular momentum carried by quarks in a transversely polarized nucleon and the generalized parton distributions $H$ and $E$, which is closely analogous to Ji's relation for the longitudinal component of the quark angular momentum in a longitudinally polarized nucleon. Neither relation is a genuine sum rule, but both offer interesting,  and complementary, possibilities for testing models and lattice calculations, but it should not be forgotten that all the terms in these equations depend on the renormalization scale $\mu$.
\bibliography{Elliot_General}

\end{document}